\documentstyle[preprint,tighten,eqsecnum,aps,prb]{revtex}
\begin{document}
\preprint{cond-mat/9310066}
\draft
\title{Exact solution for the distribution of transmission eigenvalues in a
disordered wire and comparison with random-matrix theory}
\author{C. W. J. Beenakker and B. Rejaei}
\address{Instituut-Lorentz, University of Leiden\\
P.O. Box 9506, 2300 RA Leiden, The Netherlands}
\date{October 1993}
\maketitle
\begin{abstract}
We consider the complete probability distribution $P(\{T_{n}\})$ of the
transmission eigenvalues $T_{1},T_{2},\ldots T_{N}$ of a disordered
quasi-one-dimensional conductor (length $L$ much greater than width $W$ and
mean free path $l$). The Fokker-Planck equation which describes the evolution
of $P$ with increasing $L$ is mapped onto a Schr\"{o}dinger equation by a
Sutherland-type transformation. In the absence of time-reversal symmetry (e.g.\
because of a magnetic field), the mapping is onto a free-fermion problem, which
we solve exactly. The resulting distribution is compared with the predictions
of random-matrix theory (RMT) in the metallic regime ($L\ll Nl$) and in the
insulating regime ($L\gg Nl$). We find that the logarithmic eigenvalue
repulsion of RMT is exact for $T_{n}$'s close to unity, but overestimates the
repulsion for weakly transmitting channels. The non-logarithmic repulsion
resolves several long-standing discrepancies between RMT and microscopic
theory, notably in the magnitude of the universal conductance fluctuations in
the metallic regime, and in the width of the log-normal conductance
distribution in the insulating regime.
\end{abstract}
\pacs{PACS numbers: 72.10.Bg, 05.60.+w, 72.15.Rn, 73.50.Bk}
%\newpage
\narrowtext
\section{Introduction}
\label{intro}
A fundamental problem of mesoscopic physics is to find the statistical
distribution of the scattering matrix in an ensemble of disordered conductors.
Once this is known, one can compute all moments of the conductance, and of any
other transport property, at temperatures which are sufficiently low that the
conductor is fully phase-coherent. Random-matrix theory (RMT) addresses this
problem on the basis of the assumption that all correlations between the
transmission eigenvalues are due to the jacobian from matrix to eigenvalue
space.\cite{Imr86,Mut87,Sto91} The transmission eigenvalues $T_{1},T_{2},\ldots
T_{N}$ are the eigenvalues of the matrix product $tt^{\dagger}$, where $t$ is
the $N\times N$ transmission matrix of the conductor. The jacobian is
\begin{equation}
J(\{\lambda_{n}\})=\prod_{i<j}|\lambda_{j}-\lambda_{i}|^{\beta},
\label{jacobian}
\end{equation}
where $\lambda_{i}\equiv (1-T_{i})/T_{i}$ is the ratio of reflection to
transmission probabilities ($\lambda\geq 0$, since $0\leq T\leq 1$), and
$\beta\in\{1,2,4\}$ is the symmetry index of the ensemble of scattering
matrices. [In the absence of time-reversal symmetry, one has $\beta=2$; In the
presence of time-reversal symmetry, one has $\beta=1\,(4)$ in the presence
(absence) of spin-rotation symmetry.]

If all correlations are due to the jacobian, then the probability distribution
$P(\lambda_{1},\lambda_{2},\ldots\lambda_{N})$ of the $\lambda$'s should have
the form $P\propto J\prod_{i}f(\lambda_{i})$, or equivalently,
\begin{mathletters}
\label{Pglobal}
\begin{eqnarray}
P(\{\lambda_{n}\})&=&C\exp\Bigl[-\beta\Bigl(\sum_{i<j}
u(\lambda_{i},\lambda_{j})+\sum_{i}V(\lambda_{i})\Bigr)\Bigr],
\label{Pglobala}\\
u(\lambda_{i},\lambda_{j})&=&-\ln|\lambda_{j}-\lambda_{i}|,
\label{Pglobalb}
\end{eqnarray}
\end{mathletters}%
with $V=-\beta^{-1}\ln f$ and $C$ a normalization constant. Eq.\
(\ref{Pglobal}) has the form of a Gibbs distribution at temperature
$\beta^{-1}$ for a fictitious system of classical particles moving in one
dimension in an external potential $V$, with a logarithmically repulsive
interaction $u$. All microscopic parameters (sample length $L$, width $W$, mean
free path $l$, Fermi wave length $\lambda_{\rm F}$) are contained in the single
function $V(\lambda)$. The logarithmic repulsion is independent of microscopic
parameters, because of its geometric origin.

The RMT probability distribution (\ref{Pglobal}), due to Muttalib, Pichard, and
Stone, was justified by a maximum-entropy principle for quasi-one-dimensional
(quasi-1D) conductors.\cite{Mut87,Sto91} Quasi-1D means $L\gg W$. In this limit
one can assume that the distribution of scattering matrices is only a function
of the transmission eigenvalues (isotropy assumption). The distribution
(\ref{Pglobal}) then maximizes the information entropy subject to the
constraint of a given density of eigenvalues. The function $V(\lambda)$ is
determined by this constraint and is not specified by RMT.

It was initially believed that Eq.\ (\ref{Pglobal}) would provide an exact
description in the quasi-1D limit, if only $V(\lambda)$ were suitably
chosen.\cite{Sto91,Mel89} However, it was shown recently by one of
us\cite{Bee93} that RMT is not exact, even in the quasi-1D limit. If one
computes from Eq.\ (\ref{Pglobal}) in the metallic regime the variance ${\rm
Var}\,G$ of the conductance $G=G_{0}\sum_{n}T_{n}$ (with $G_{0}=2e^{2}/h$), one
finds\cite{Bee93}
\begin{equation}
{\rm Var\,}G/G_{0}=\frac{1}{8}\beta^{-1},\label{UCFglobal}
\end{equation}
independent of the form of $V(\lambda)$. The diagrammatic perturbation
theory\cite{Alt85,Lee85} of universal conductance fluctuations (UCF) gives
instead
\begin{equation}
{\rm Var\,}G/G_{0}=\frac{2}{15}\beta^{-1}\label{UCF}
\end{equation}
for a quasi-1D conductor. The difference between the coefficients $\frac{1}{8}$
and $\frac{2}{15}$ is tiny, but it has the fundamental implication that the
interaction between the $\lambda$-variables is not precisely logarithmic, or in
other words, that there exist correlations between the transmission eigenvalues
over and above those induced by the jacobian.

What then is the status of the random-matrix theory of quantum transport? It is
obviously highly accurate, so that the true eigenvalue interaction should be
close to logarithmic. Is there perhaps a cutoff for large separation of the
$\lambda$'s? Or is the true interaction a many-body interaction, which can not
be reduced to the sum of pairwise interactions? That is the problem addressed
in this paper. A brief account of our results was reported in a recent
Letter.\cite{Rej94}

The transport problem considered here has a counterpart in equilibrium. The
Wigner-Dyson RMT of the statistics of the eigenvalues $\{E_{n}\}$ of a random
hamiltonian yields a probability distribution of the form (\ref{Pglobal}), with
a logarithmic repulsion between the energy levels.\cite{Meh67} It was shown by
Efetov\cite{Efe83} and by Al'tshuler and Shklovski\u{\i}\cite{Alt86} that the
logarithmic level repulsion in a small disordered particle (diameter $L$,
diffusion constant $D$) holds for energy separations small compared to the
Thouless energy $E_{\rm c}\equiv\hbar D/L^{2}$. For larger separations the
interaction potential decays algebraically.\cite{Jal93} As we will see, the way
in which the RMT of quantum transport breaks down is quite different: The
interaction $u(\lambda_{i},\lambda_{j})=-\ln|\lambda_{j}-\lambda_{i}|$ is exact
for $\lambda_{i},\lambda_{j}\ll 1$, i.e.\ for strongly transmitting scattering
channels (recall that $\lambda\ll 1$ implies $T\equiv(1+\lambda)^{-1}$ close to
unity). For weakly transmitting channels the repulsion is still logarithmic,
but reduced by a factor of two from what one would expect from the jacobian.
This modified interaction explains the $\frac{1}{8}$~---~$\frac{2}{15}$
discrepancy in the UCF in the metallic regime,\cite{Bee93} and it also explains
a missing factor of two in the width of the log-normal distribution of the
conductance in the insulating regime.\cite{Pic91}

Our analysis is based on the Dorokhov-Mello-Pereyra-Kumar (DMPK) equation
\begin{equation}
l\frac{\partial P}{\partial L}=
\frac{2}{\beta N+2-\beta}\sum_{i=1}^{N}
\frac{\partial}{\partial\lambda_{i}}\lambda_{i}(1+\lambda_{i})
J\frac{\partial}{\partial\lambda_{i}}J^{-1}P,
\label{DMPK}
\end{equation}
with ballistic initial condition $\lim_{L\rightarrow
0}P=\prod_{i}\delta(\lambda_{i}-0^{+})$, which describes the evolution of the
eigenvalue distribution function in an ensemble of disordered wires of
increasing length. Eq.\ (\ref{DMPK}) was derived by Dorokhov,\cite{Dor82} (for
$\beta=2$) and by Mello, Pereyra, and Kumar,\cite{Mel88} (for $\beta=1$, with
generalizations to $\beta=2,4$ in Refs.\ \onlinecite{Mel91,Mac92}) by computing
the incremental change of the transmission eigenvalues upon attachment of a
thin slice to the wire. It is assumed that the conductor is weakly disordered,
$l\gg\lambda_{\rm F}$, so that the scattering in the thin slice can be treated
by perturbation theory. A key simplification is the isotropy assumption that
the flux incident in one scattering channel is, on average, equally distributed
among all outgoing channels. This assumption restricts the applicability of the
DMPK equation to the quasi-1D regime $L\gg W$, since it ignores the finite time
scale for transverse diffusion.

Eq.\ (\ref{DMPK}) has the form of a diffusion equation in a complicated
$N$-dimensional space (identified as a certain Riemannian manifold in Ref.\
\onlinecite{Huf90}). For a coordinate-free ``supersymmetry formulation'' of
this diffusion process, see Refs.\ \onlinecite{Iid90,Zir92}. The similarity to
diffusion in real space has been given further substance by the
demonstration\cite{Mel88c} that Eq.\ (\ref{DMPK}) holds on length scales $\gg
l$ regardless of the microscopic scattering properties of the conductor.

The diffusion equation (\ref{DMPK}) has been studied extensively for more than
ten years. Exact solutions have been obtained by Mel'nikov\cite{Mel81} and
Mello\cite{Mel86} for the case $N=1$ of a single degree of freedom (when
$J\equiv 1$). For $N>1$ the strong coupling of the scattering channels by the
jacobian (\ref{jacobian}) prevented an exact solution by standard methods. The
problem simplifies drastically deep in the localized regime ($L\gg Nl$), when
the scattering channels become effectively decoupled. Pichard\cite{Pic91} has
computed from Eq.\ (\ref{DMPK}) the log-normal distribution of the conductance
in this regime, and has found an excellent agreement with numerical simulations
of a quasi-1D Anderson insulator. In the metallic regime ($L\ll Nl$), Mello and
Stone\cite{Mel91,Mel88b} were able to compute the first two moments of the
conductance, in precise agreement with the diagrammatic perturbation theory of
weak localization and UCF [Eq.\ (\ref{UCF})] in the quasi-1D limit. (Their
method of moments has also been applied to the shot noise,\cite{Jon92} where
there is no diagrammatic theory to compare with.) More general calculations of
the weak localization effect\cite{Bee94} and of universal
fluctuations\cite{Cha93} [for arbitrary transport properties of the form
$A=\sum_{n}a(T_{n})$] were recently developed, based on linearization of Eq.\
(\ref{DMPK}) in the fluctuations of the $\lambda$'s around their mean positions
(valid in the large-$N$ metallic regime, when the fluctuations are small). The
work of Chalker and Mac\^{e}do\cite{Cha93} was motivated by the same
$\frac{1}{8}$~---~$\frac{2}{15}$ discrepancy\cite{Bee93} as the present paper
and Ref.\ \onlinecite{Rej94}, with which it has some overlap.

None of these calculations suffices to determine the form of the eigenvalue
interaction, which requires knowledge of the complete distribution function.
Here we wish to present (in considerable more detail than in our
Letter\cite{Rej94}) the exact solution of Eq.\ (\ref{DMPK}) for $\beta =2$.

The outline of this paper is as follows. In Sec. II we solve Eq.\ (\ref{DMPK})
exactly, for all $N$ and $L$, for the case $\beta=2$. The method of solution is
a mapping onto a model of non-interacting fermions, inspired by Sutherland's
mapping of a different diffusion equation.\cite{Sut72} The case $\beta=2$ is
special, because for other values of $\beta$ the mapping introduces
interactions between the fermions. The free-fermion problem, which is obtained
for $\beta=2$, has the character of a one-dimensional scattering problem in
imaginary time. The absence of a ground state is a significant complication,
compared with Sutherland's problem.\cite{Sim93,Muc93,Rej93} The exact solution
which we obtain has the form of a determinant of an $N\times N$ matrix. The
determinant can be evaluated in closed form in the metallic regime $L\ll Nl$
and in the insulating regime $L\gg Nl$. These two opposite regimes are
discussed separately in Secs.\ III and IV. We conclude in Sec.\ V with a
comparison of the solution of Eq.\ (\ref{DMPK}) with the probability
distribution (\ref{Pglobal}) of random-matrix theory.

\section{Exact solution}
\label{exact}
The solution of the Dorokhov-Mello-Pereyra-Kumar (DMPK) equation (\ref{DMPK})
proceeds in a series of steps, which we describe in separate subsections.

\subsection{Transformation of variables}
\label{transformation}
The DMPK equation (\ref{DMPK}) can be written in the form of an $N$-dimensional
Fokker-Planck equation,
\begin{mathletters}
\label{FPlambda}
\begin{eqnarray}
&&\frac{\partial}{\partial s}P(\{\lambda_{n}\},s)=
\sum_{i=1}^{N}\frac{\partial}{\partial\lambda_{i}}D(\lambda_{i})
\left(\frac{\partial P}{\partial\lambda_{i}}+
\beta P\frac{\partial}{\partial\lambda_{i}}\Omega(\{\lambda_{n}\})
\right),\label{FPlambda_a}\\
&&D(\lambda)=\frac{2}{\gamma}\,\lambda(1+\lambda),\label{FPlambdab}\\
&&\Omega(\{\lambda_{n}\})=-\sum_{i<j}\ln|\lambda_{j}-\lambda_{i}|,
\label{FPlambdac}
\end{eqnarray}
\end{mathletters}%
where we have abbreviated $s=L/l$, $\gamma=\beta N+2-\beta$. Eq.\
(\ref{FPlambda}) is the diffusion equation in ``time'' $s$ of a one-dimensional
gas of $N$ classical particles with a logarithmically repulsive interaction
potential $\Omega$. The diffusion takes place at temperature $\beta^{-1}$ in a
fictitious non-uniform viscous fluid with diffusion coefficient $D(\lambda)$.

The position dependence of the diffusion coefficient is problematic. We seek to
eliminate it by a transformation of variables. Let $\{x_{n}\}$ be a new set of
$N$ independent variables, related to the $\lambda$'s by
$\lambda_{n}=f(x_{n})$. The new probability distribution
$P(\{x_{n}\},s)=P(\{\lambda_{n}\},s)\prod_{i}|f'(x_{i})|$ still satisfies a
Fokker-Planck equation, but with a new potential $\Omega(\{x_{n}\})$ and a new
diffusion coefficient $D(x)$. The potential transforms as
$\Omega\rightarrow\Omega-\beta^{-1}\sum_{i}\ln|f'(x_{i})|$, while the diffusion
coefficient transforms as  $D\rightarrow D/f'(x)^{2}$. In order to obtain an
$x$-independent diffusion coefficient, we thus need to choose $f(x)$ such that
$f(x)[1+f(x)]/f'(x)^{2}={\rm constant}$. The choice $f(x)=\sinh^{2}x$ does it.

We therefore transform to a new set of variables $\{x_{n}\}$, defined by
\begin{equation}
\lambda_{n}=\sinh^{2}x_{n},\;\;T_{n}=1/\cosh^{2}x_{n}.
\label{Txdef}
\end{equation}
Since $T_{n}\in[0,1]$, $x_{n}\geq 0$. The probability distribution of the
$x$-variables satisfies a Fokker-Planck equation with {\em constant\/}
diffusion coefficient,
\begin{mathletters}
\label{FokkerPlanck}
\begin{eqnarray}
&&\frac{\partial}{\partial s}P(\{x_{n}\},s)=
\frac{1}{2\gamma}\sum_{i=1}^{N}\frac{\partial}{\partial x_{i}}
\left(\frac{\partial P}{\partial x_{i}}+
\beta P\frac{\partial}{\partial x_{i}}\Omega(\{x_{n}\})\right),
\label{FPxa}\\
&&\Omega(\{x_{n}\})=-\sum_{i<j}\ln|\sinh^{2}x_{j}-\sinh^{2}x_{i}|
-\frac{1}{\beta}\sum_{i}\ln|\sinh 2x_{i}|.\label{FPxb}
\end{eqnarray}
\end{mathletters}%
It turns out that the $x$-variables have a special physical significance: The
ratio $L/x_{n}$ equals the channel-dependent localization length of the
conductor.\cite{Sto91}

\subsection{From Fokker-Planck to Schr\"{o}dinger equation}
\label{FPtoS}
Sutherland\cite{Sut72} has shown that a Fokker-Planck equation with constant
diffusion coefficient and with a logarithmic interaction potential can be
mapped onto a Schr\"{o}dinger equation with an inverse-square interaction which
{\em vanishes\/} for $\beta=2$. The Fokker-Planck equation (\ref{FokkerPlanck})
does have a constant diffusion coefficient, but the interaction is not
logarithmic. It is not obvious that Sutherland's mapping onto a free-fermion
problem should work for the non-translationally invariant interaction
(\ref{FPxb}), but surprisingly enough it does.

To map the Fokker-Planck equation (\ref{FokkerPlanck}) onto a Schr\"{o}dinger
equation we substitute
\begin{equation}
P(\{x_{n}\},s)=\exp\left[-\case{1}{2}\beta
\Omega(\{x_{n}\})\right]\Psi(\{x_{n}\},s).\label{subs}
\end{equation}
This is a variation on Sutherland's transformation,\cite{Sut72} which we used
in Ref.\ \onlinecite{Rej93} in a different context. Substitution of Eq.\
(\ref{subs}) into Eq.\ (\ref{FPxa}) yields for $\Psi$ the equation
\begin{equation}
-\frac{\partial\Psi}{\partial s}=-\frac{1}{2\gamma}
\sum_{i=1}^{N}\frac{\partial^{2}\Psi}{\partial x_{i}^{2}}+
\frac{\beta}{4\gamma}\Psi\sum_{i=1}^{N}\left[\frac{\beta}{2}
\left(\frac{\partial\Omega}{\partial x_{i}}\right)^{2}-
\frac{\partial^{2}\Omega}{\partial x_{i}^{2}}\right].\label{Schro1}
\end{equation}
The expression between square brackets is evaluated as follows (we abbreviate
$\xi_{i}=\cosh 2x_{i}$):
\begin{eqnarray}
\sum_{i=1}^{N}\frac{\partial^{2}\Omega}{\partial x_{i}^{2}}&=&
4\sum_{i}\sum_{j(\neq i)}\left(
\frac{\xi^{2}_{i}-1}{(\xi_{j}-\xi_{i})^{2}}+
\frac{\xi_{i}}{\xi_{j}-\xi_{i}}\right)+
\frac{4}{\beta}\sum_{i}\frac{1}{\xi^{2}_{i}-1}\nonumber\\
&=&4\sum_{i}\sum_{j(\neq i)}\frac{\xi^{2}_{i}-1}{(\xi_{j}-\xi_{i})^{2}}+
\frac{4}{\beta}\sum_{i}\frac{1}{\xi^{2}_{i}-1}
-4{N\choose 2},\label{expand1}\\
\sum_{i=1}^{N}\left(\frac{\partial\Omega}{\partial x_{i}}\right)^{2}&=&
4\sum_{i}\sum_{j(\neq i)}\sum_{k(\neq i)}
\frac{\xi^{2}_{i}-1}{(\xi_{j}-\xi_{i})(\xi_{k}-\xi_{i})}
-\frac{8}{\beta}\sum_{i}\sum_{j(\neq i)}\frac{\xi_{i}}{\xi_{j}-\xi_{i}}
+\frac{4}{\beta^{2}}\sum_{i}\frac{\xi^{2}_{i}}{\xi^{2}_{i}-1}\nonumber\\
&=&4\sum_{i}\sum_{j(\neq i)}\frac{\xi^{2}_{i}-1}{(\xi_{j}-\xi_{i})^{2}}
+\frac{4}{\beta^{2}}\sum_{i}\frac{1}{\xi^{2}_{i}-1}
+\frac{4}{\beta^{2}}N+\frac{8}{\beta}{N\choose 2}
+8{N\choose 3}.\label{expand2}
\end{eqnarray}
In the final equality we have used that for any three distinct indices $i,j,k$
\begin{equation}
\frac{\xi^{2}_{i}-1}{(\xi_{j}-\xi_{i})(\xi_{k}-\xi_{i})}+
\frac{\xi^{2}_{j}-1}{(\xi_{i}-\xi_{j})(\xi_{k}-\xi_{j})}+
\frac{\xi^{2}_{k}-1}{(\xi_{i}-\xi_{k})(\xi_{j}-\xi_{k})}\equiv 1,
\label{expand3}
\end{equation}
so that the triple sum over $k\neq i\neq j$ collapses to a double sum over
$i\neq j$. Collecting results, we find that $\Psi$ satisfies a Schr\"{o}dinger
equation in imaginary time,
\begin{mathletters}
\label{Schrodinger}
\begin{eqnarray}
-\frac{\partial\Psi}{\partial s}&=&({\cal H}-U)\Psi,
\label{Schrodingera}\\
{\cal H}&=&-\frac{1}{2\gamma}\sum_{i}\left(\frac{\partial^{2}}
{\partial x_{i}^{2}}+\frac{1}{\sinh^{2}2x_{i}}\right)
+\frac{\beta(\beta-2)}{2\gamma}\sum_{i<j}
\frac{\sinh^{2}2x_{j}+\sinh^{2}2x_{i}}
{(\cosh 2x_{j}-\cosh 2x_{i})^{2}},\label{Hdef}\\
U&=&-\frac{N}{2\gamma}-N(N-1)\frac{\beta}{\gamma}-
N(N-1)(N-2)\frac{\beta^{2}}{6\gamma}.\label{Udef}
\end{eqnarray}
\end{mathletters}%

The interaction potential in the hamiltonian (\ref{Hdef}) is attractive for
$\beta=1$ and repulsive for $\beta=4$. For $\beta=2$ the interaction vanishes
identically,  reducing ${\cal H}$ to a sum of single-particle hamiltonians
${\cal H}_{0}$,
\begin{equation}
{\cal H}_{0}=-\frac{1}{4N}\frac{\partial^{2}}
{\partial x^{2}}-\frac{1}{4N\sinh^{2}2x}.\label{H0def}
\end{equation}
(Note that $\gamma=2N$ for $\beta=2$.) It might be possible to solve also the
interacting Schr\"{o}\-dinger equation (\ref{Schrodinger}) for $\beta=1$ or 4,
by some modification of techniques developed for the Suther\-land
hamil\-tonian,\cite{Sut72,Sim93,Muc93} but in this paper we focus on the
simplest case $\beta=2$ of broken time-reversal symmetry.

To complete the mapping onto a single-particle problem, we need to consider the
boundary condition at the edge $x=0$. (Recall that $x\geq 0$.) Conservation of
probability implies for $P$ the boundary condition (one for each $i=1,2,\ldots
N$)
\begin{equation}
\lim_{x_{i}\rightarrow 0}\left(\frac{\partial P}{\partial x_{i}}
+\beta P\frac{\partial\Omega}{\partial x_{i}}\right)=0.
\label{Pboundary}
\end{equation}
According to Eq.\ (\ref{subs}), the corresponding boundary condition on $\Psi$
is
\begin{equation}
\lim_{x_{i}\rightarrow 0}\left(\frac{\partial\Psi}{\partial x_{i}}
+\case{1}{2}\beta\Psi\frac{\partial\Omega}{\partial x_{i}}\right)=0,
\label{Psiboundary1}
\end{equation}
which in view of Eq.\ (\ref{FPxb}) simplifies to
\begin{equation}
\lim_{x_{i}\rightarrow 0}\left(\frac{\partial\Psi}{\partial x_{i}}
-\frac{\Psi}{\sinh 2x_{i}}\right)=0,
\label{Psiboundary2}
\end{equation}
independent of $\beta$. Fortunately, the boundary condition does not couple
different degrees of freedom, so that we have indeed obtained a single-particle
problem for $\beta=2$.

\subsection{From probability distribution to fermion Green's function}

We seek a solution $P(\{x_{n}\},s\,|\,\{y_{n}\})$ of the Fokker-Planck equation
(\ref{FokkerPlanck}) with symmetrized delta-function initial condition
\begin{equation}
P(\{x_{n}\},0\,|\,\{y_{n}\})=\frac{1}{N!}\sum_{\pi}\,
\prod_{i=1}^{N}\delta(x_{i}-y_{\pi_{i}}).\label{Pinitial}
\end{equation}
The sum in Eq.\ (\ref{Pinitial}) is over all $N!$ permutations of $1,2,\ldots
N$. Eventually, we will take the limit $\{y_{n}\}\rightarrow 0$ of a ballistic
initial condition, but it is convenient to first consider the more general
initial condition (\ref{Pinitial}). In this subsection we use the mapping onto
a Schr\"{o}dinger equation of the previous subsection to relate the probability
distribution $P(\{x_{n}\},s\,|\,\{y_{n}\})$ to the $N$-fermion Green's function
$G(\{x_{n}\},s\,|\,\{y_{n}\})$.

We first note that, since $\exp(-\beta\Omega)$ is an $s$-independent solution
of the Fokker-Planck equation (\ref{FokkerPlanck}),
$\exp(-\frac{1}{2}\beta\Omega)$ is an $s$-independent solution of the
Schr\"{o}dinger equation (\ref{Schrodinger}) [in view of the mapping
(\ref{subs})]. For a particular ordering of the $x_{n}$'s, the function
$\Psi_{0}\propto\exp(-\frac{1}{2}\beta\Omega)$ is therefore an eigenfunction of
the $N$-fermion hamiltonian ${\cal H}$ with eigenvalue $U$. Anti-symmetrization
yields the fermion eigenstate
\begin{equation}
\Psi_{0}(\{x_{n}\})=C\exp\left[-\case{1}{2}\beta\Omega(\{x_{n}\})\right]
\prod_{i<j}\frac{x_{j}-x_{i}}{|x_{j}-x_{i}|},\label{Psi0}
\end{equation}
with $C$ a normalization constant.

We obtain the $N$-fermion Green's function $G$ from the probability
distribution $P$ by the similarity transformation
\begin{equation}
G(\{x_{n}\},s\,|\,\{y_{n}\})=
\Psi_{0}^{-1}(\{x_{n}\})P(\{x_{n}\},s\,|\,\{y_{n}\})
\Psi_{0}^{\vphantom{-1}}(\{y_{n}\}).\label{PGrelation}
\end{equation}
To verify this, we first observe that $G$ is by construction anti-symmetric
under a permutation of two $x$ or two $y$ variables. For a given order of the
$x_{n}$'s, the function $G$ satisfies the Schr\"{o}dinger equation
\begin{equation}
-\frac{\partial G}{\partial s}=({\cal H}-U)G,
\label{SchrodingerG}
\end{equation}
in view of Eqs.\ (\ref{subs}), (\ref{Schrodingera}), and (\ref{Psi0}). Finally,
Eq.\ (\ref{Pinitial}) implies the initial condition
\begin{equation}
G(\{x_{n}\},0\,|\,\{y_{n}\})=\frac{1}{N!}\sum_{\pi}\sigma_{\pi}
\prod_{i=1}^{N}\delta(x_{i}-y_{\pi_{i}}),
\label{Ginitial}
\end{equation}
with $\sigma_{\pi}$ the sign of the permutation. Hence $G$ is indeed the
$N$-fermion Green's function.

The relation (\ref{PGrelation}) holds for any $\beta$. In the remainder of this
paper we consider the non-interacting case $\beta=2$. The eigenstate
(\ref{Psi0}) then takes the form
\begin{equation}
\Psi_{0}(\{x_{n}\})=C\prod_{i<j}(\sinh^{2}x_{j}-\sinh^{2}x_{i})
\prod_{i}(\sinh 2x_{i})^{1/2}.\label{Psi0beta2}
\end{equation}
The $N$-fermion Green's function $G$ becomes a Slater determinant of the
single-particle Green's function $G_{0}$,
\begin{equation}
G(\{x_{n}\},s\,|\, \{y_{n}\})=\frac{{\rm e}^{Us}}{N!}\,{\rm
Det\,}G_{0}(x_{n},s\,|\,y_{m}),\label{Gdef}
\end{equation}
where ${\rm Det\,}a_{nm}$ denotes the determinant of the $N\times N$ matrix
with elements $a_{nm}$. The function $G_{0}(x,s\,|\, y)$ is a solution of the
single-particle Schr\"{o}dinger equation $-\partial G_{0}/\partial s={\cal
H}_{0}G_{0}$ in the variable $x$, with initial condition $G(x,0\,|\,
y)=\delta(x-y)$. In the following subsection we will compute the
single-particle Green's function $G_{0}$. The probability distribution $P$, for
$\beta=2$, then follows from Eqs.\ (\ref{PGrelation}), (\ref{Psi0beta2}), and
(\ref{Gdef}):
\begin{equation}
P(\{x_{n}\},s\,|\,\{y_{n}\})=\frac
{\prod_{i<j}(\sinh^{2}x_{j}-\sinh^{2}x_{i})\prod_{i}(\sinh 2x_{i})^{1/2}}
{\prod_{i<j}(\sinh^{2}y_{j}-\sinh^{2}y_{i})\prod_{i}(\sinh 2y_{i})^{1/2}}
\,\frac{{\rm e}^{Us}}{N!}\,
{\rm Det\,}G_{0}(x_{n},s\,|\,y_{m}).\label{PGrelationbeta2}
\end{equation}

\subsection{Computation of Green's function}
\label{Green}

To compute the Green's function $G_{0}$ of the single-particle hamiltonian
(\ref{H0def}) we need to solve the eigenvalue equation
\begin{equation}
-\frac{1}{4N}\,\frac{d^{\,2}}{d x^{2}}\psi(x)
-\frac{1}{4N}\,\frac{\psi(x)}{\sinh^{2}2x}=
\varepsilon\psi(x),\label{eigenvalue1}
\end{equation}
with the boundary condition dictated by Eq.\ (\ref{Psiboundary2}),
\begin{equation}
\lim_{x\rightarrow 0}\left(\frac{d\psi}{dx}
-\frac{\psi}{\sinh 2x}\right)=0.\label{psiboundary}
\end{equation}
We have found that the substitution
\begin{equation}
\psi(x)=(\sinh 2x)^{1/2}\,f(\cosh 2x)\label{psifdef}
\end{equation}
transforms Eq.\ (\ref{eigenvalue1}) into Legendre's differential equation in
the variable $z=\cosh 2x$,
\begin{equation}
\frac{d}{dz}\left[(1-z^{2})\frac{d}{dz}f(z)\right]=
(N\varepsilon+\case{1}{4})f(z).\label{Legendre}
\end{equation}
The boundary condition (\ref{psiboundary}) restricts the solutions of Eq.\
(\ref{Legendre}) to the Legendre functions of the first kind ${\rm
P}_{\nu}(z)$. The index $\nu$ is given by $\nu=-\frac{1}{2}+\frac{1}{2}{\rm
i}k$ with $k$ a real number. (These Legendre functions are also known as
``toroidal functions'', because they appear as solutions to the Laplace
equation in toroidal coordinates.) The numbers $\nu$, $k$, and $\varepsilon$
are related by $-\nu(\nu+1)=N\varepsilon+\case{1}{4}$ and
$\varepsilon=\frac{1}{4}k^{2}/N$. We can restrict ourselves to $k\geq 0$, since
the functions ${\rm P}_{-\frac{1}{2}+\frac{1}{2}{\rm i}k}$ and ${\rm
P}_{-\frac{1}{2}-\frac{1}{2}{\rm i}k}$ are identical.

We conclude that the spectrum of ${\cal H}_{0}$ is continuous, with positive
eigenvalues $\varepsilon=\frac{1}{4}k^{2}/N$. The eigenfunctions $\psi_{k}(x)$
are real functions given by
\begin{equation}
\psi_{k}(x)=[\pi k\tanh(\case{1}{2}\pi k)\sinh(2x)]^{1/2}\,{\rm
P}_{\frac{1}{2}({\rm i}k-1)}(\cosh 2x).\label{psikdef}
\end{equation}
They form a complete and orthonormal set,
\begin{eqnarray}
\int_{0}^{\infty}\!\!dk\,\psi_{k}(x)\psi_{k}(x')&=&2\pi\delta(x-x'),
\label{completepsi}\\
\int_{0}^{\infty}\!\!dx\,\psi_{k}(x)\psi_{k'}(x)&=&2\pi\delta(k-k'),
\label{normalizepsi}
\end{eqnarray}
in accordance with the inversion formula in Ref.\ \onlinecite{Erdelyi}. The
single-particle Green's function $G_{0}$ has the corresponding spectral
representation
\begin{eqnarray}
G_{0}(x,s\,|\, y)&=&
(2\pi)^{-1}\int_{0}^{\infty}\!\!dk\,\exp(-\case{1}{4}k^{2}s/N)\,
\psi_{k}(x)\psi_{k}(y)\nonumber\\
&=&\case{1}{2}(\sinh 2x\sinh 2y)^{1/2}\int_{0}^{\infty}\!\!dk\,
\exp(-\case{1}{4}k^{2}s/N)k\tanh(\case{1}{2}\pi k)\nonumber\\
&&\hspace{3cm}\mbox{}\times{\rm P}_{\frac{1}{2}({\rm i}k-1)}(\cosh 2x)
{\rm P}_{\frac{1}{2}({\rm i}k-1)}(\cosh 2y).\label{G0def}
\end{eqnarray}

\subsection{Ballistic initial condition}
\label{ballistic}

Eqs.\ (\ref{PGrelationbeta2}) and (\ref{G0def}) together determine the
probability distribution $P(\{x_{n}\},s\,|\,\{y_{n}\})$ with initial condition
(\ref{Pinitial}),
\begin{eqnarray}
P&=&C(s)\frac
{\prod_{i<j}(\sinh^{2}x_{j}-\sinh^{2}x_{i})\prod_{i}(\sinh 2x_{i})}
{\prod_{i<j}(\sinh^{2}y_{j}-\sinh^{2}y_{i})}\nonumber\\
&&\mbox{}\times{\rm Det}\left[\int_{0}^{\infty}\!\!dk\,
\exp(-\case{1}{4}k^{2}s/N)k\tanh(\case{1}{2}\pi k)
{\rm P}_{\frac{1}{2}({\rm i}k-1)}(\cosh 2x_{n})
{\rm P}_{\frac{1}{2}({\rm i}k-1)}(\cosh 2y_{m})\right]\nonumber\\
&=&C(s)\prod_{i<j}(\sinh^{2}x_{j}-\sinh^{2}x_{i})\prod_{i}(\sinh 2x_{i})
\int_{0}^{\infty}\!\!dk_{1}\int_{0}^{\infty}\!\!dk_{2}
\cdots\int_{0}^{\infty}\!\!dk_{N}\nonumber\\
&&\mbox{}\times
\prod_{i}\left[\exp(-\case{1}{4}k^{2}_{i}s/N)k_{i}
\tanh(\case{1}{2}\pi k_{i})
{\rm P}_{\frac{1}{2}({\rm i}k_{i}-1)}(\cosh 2x_{i})\right]
\frac{
{\rm Det\,}{\rm P}_{\frac{1}{2}({\rm i}k_{n}-1)}(\cosh 2y_{m})}
{\prod_{i<j}(\sinh^{2}y_{j}-\sinh^{2}y_{i})}.
\label{Pgeneral}
\end{eqnarray}
We have absorbed all $x$ and $y$ independent factors into the function $C(s)$,
which is fixed by the requirement that $P$ is normalized to unity,
\begin{equation}
\int_{0}^{\infty}\!\!dx_{1}\int_{0}^{\infty}\!\!dx_{2}
\cdots\int_{0}^{\infty}\!\!dx_{N}\,P=1.
\label{normalizeP}
\end{equation}
In the second equality in Eq.\ (\ref{Pgeneral}) we have applied the identity
${\rm Det\,}(b_{n}a_{nm})=(\prod_{i}b_{i}){\rm Det\,}a_{nm}$ to isolate the
factors containing the $y$-variables.

Now it remains to take the limit $\{y_{n}\}\rightarrow 0$ of a ballistic
initial condition. The limit is tricky because it involves a cancellation of
zeroes of the determinant in the numerator with zeroes of the alternating
function in the denominator. It is convenient to first write the alternating
function as a Vandermonde determinant,
\begin{equation}
\prod_{i<j}(\sinh^{2}y_{j}-\sinh^{2}y_{i})=
{\rm Det\,}(\sinh^{2}y_{m})^{n-1}.
\label{sinhvdm}
\end{equation}
Next, we expand the Legendre function in powers of $\sinh^{2}y$,
\begin{equation}
{\rm P}_{\frac{1}{2}({\rm i}k-1)}(\cosh 2y)=\sum_{p=1}^{\infty}c_{p}(k)
(\sinh^{2}y)^{p-1}.\label{Pexpansion}
\end{equation}
The factors $c_{p}(k)$ are polynomials in $k^2$, with $c_{1}(k)\equiv 1$ and
\begin{equation}
c_{p}(k)=(-1)^{p-1}[2^{p-1}(p-1)!]^{-2}\,(k^{2}+1^{2})(k^{2}+3^2)
\cdots(k^{2}+(2p-3)^{2})\label{cpkdef}
\end{equation}
for $p\geq 2$. In the limit $y\rightarrow 0$, we can truncate the expansion
(\ref{Pexpansion}) after the first $N$ terms, that is to say,
\begin{equation}
\lim_{\{y_{n}\}\rightarrow 0}\frac{
{\rm Det\,}{\rm P}_{\frac{1}{2}({\rm i}k_{n}-1)}(\cosh 2y_{m})}
{\prod_{i<j}(\sinh^{2}y_{j}-\sinh^{2}y_{i})}
=\lim_{\{y_{n}\}\rightarrow 0}
\frac{{\rm Det}\left[
\sum_{p=1}^{N}c_{p}(k_{n})(\sinh^{2}y_{m})^{p-1}\right]}
{{\rm Det\,}(\sinh^{2}y_{m})^{n-1}}.\label{detquotient}
\end{equation}
The numerator on the r.h.s.\ of Eq.\ (\ref{detquotient}) factors as the product
of two determinants, one of which is just the Vandermonde determinant in the
denominator, so that the whole quotient reduces to the single determinant ${\rm
Det\,}c_{m}(k_{n})$. This determinant can be simplified by means of the
identity
\begin{equation}
{\rm Det\,}c_{m}(k_{n})=c_{0}\,{\rm Det\,}(k_{n}^{2})^{m-1},\label{cdet}
\end{equation}
with $c_{0}$ a numerical coefficient. Eq.\ (\ref{cdet}) holds because the
determinant of a matrix is unchanged if any one column of the matrix is added
to any other column, so that we can reduce the polynomial $c_{m}(k)$ in $k^{2}$
of degree $m-1$ to just its highest order term $k^{2(m-1)}$ times a numerical
coefficient.

Collecting results, we find
\begin{equation}
\lim_{\{y_{n}\}\rightarrow 0}\frac{
{\rm Det\,}{\rm P}_{\frac{1}{2}({\rm i}k_{n}-1)}(\cosh 2y_{m})}
{\prod_{i<j}(\sinh^{2}y_{j}-\sinh^{2}y_{i})}
=c_{0}\,{\rm Det\,}(k_{n}^{2})^{m-1}.\label{limit}
\end{equation}
Substituting into Eq.\ (\ref{Pgeneral}), and absorbing the coefficient $c_{0}$
in the function $C(s)$, we obtain the probability distribution $P(\{x_{n}\},s)$
for a ballistic initial condition,
\begin{eqnarray}
P(\{x_{n}\},s)&=&C(s)\prod_{i<j}(\sinh^{2}x_{j}-\sinh^{2}x_{i})
\prod_{i}(\sinh 2x_{i})\nonumber\\
&&\mbox{}\times{\rm Det}\left[\int_{0}^{\infty}\!\!dk\,
\exp(-\case{1}{4}k^{2}s/N)\tanh(\case{1}{2}\pi k)k^{2m-1}\,
{\rm P}_{\frac{1}{2}({\rm i}k-1)}(\cosh 2x_{n})\right].
\label{final}
\end{eqnarray}
This is the exact solution of the DMPK equation for the case $\beta=2$.

\section{Metallic regime}
\label{metallic}

\subsection{Probability distribution}
\label{distribution}
The solution (\ref{final}) holds for any $s$ and $N$. It can be simplified in
the regime $1\ll s\ll N$ of a conductor which is long compared to the mean free
path $l$ but short compared to the localization length $Nl$. This is the
metallic regime. The dominant contribution to the integral over $k$ in Eq.\
(\ref{final}) then comes from the range $k\gtrsim (N/s)^{1/2}\gg 1$. In this
range $\tanh(\case{1}{2}\pi k)\rightarrow 1$ and the Legendre function
simplifies to a Bessel function,\cite{Note1}
\begin{equation}
{\rm P}_{\frac{1}{2}({\rm i}k-1)}(\cosh 2x)=J_{0}(kx)
\left(\frac{2x}{\sinh 2x}\right)^{\frac{1}{2}}
\;\;{\rm for}\;k\gg 1,\;x\gg 1/k.
\label{PJrelation}
\end{equation}
The second condition $x\gg 1/k$ on Eq.\ (\ref{PJrelation}) implies the
restriction $x\gg(s/N)^{1/2}$, which is irrelevant since $s/N\ll 1$. The
$k$-integration can now be carried out analytically,
\begin{equation}
\int_{0}^{\infty}\!\!dk\,\exp(-\case{1}{4}k^{2}s/N)
k^{2m-1}\,J_{0}(kx_{n})=\case{1}{2}(m-1)!\,(4N/s)^{m}
\exp(-x_{n}^{2}N/s)L_{m-1}(x_{n}^{2}N/s),\label{integral}
\end{equation}
with $L_{m-1}$ a Laguerre polynomial. We then apply the determinantal identity
\begin{equation}
{\rm Det\,}L_{m-1}(x_{n}^{2}N/s)=c\,{\rm
Det\,}(x_{n}^{2})^{m-1}=c\prod_{i<j}(x_{j}^{2}-x_{i}^{2}),
\label{Ldet}
\end{equation}
with $c$ an $x$-independent number [which can be absorbed in $C(s)$].  Eq.\
(\ref{Ldet}) is derived in the same way as Eq.\ (\ref{cdet}), by combining
columns of the matrix of polynomials in $x^{2}$. Collecting results, we find
that the general solution (\ref{final}) simplifies in the metallic regime to
\begin{equation}
P(\{x_{n}\},s)=C(s)\prod_{i<j}\left[(\sinh^{2}x_{j}-\sinh^{2}x_{i})
(x_{j}^{2}-x_{i}^{2})\right]
\prod_{i}\left[\exp(-x_{i}^{2}N/s)(x_{i}\sinh
2x_{i})^{1/2}\right].\label{final2}
\end{equation}

In the remainder of this section we use the probability distribution
(\ref{final2}) to compute various statistical quantities of interest. For that
purpose it is convenient to write $P$ as a Gibbs distribution,
\begin{mathletters}
\label{final3x}
\begin{eqnarray}
P(\{x_{n}\},s)&=&C(s)\exp\Bigl[-\beta\Bigl
(\sum_{i<j}u(x_{i},x_{j})+\sum_{i}V(x_{i},s)\Bigr)\Bigr],
\label{final3Px}\\
u(x_{i},x_{j})&=&-\case{1}{2}\ln|\sinh^{2}x_{j}-\sinh^{2}x_{i}|
-\case{1}{2}\ln|x_{j}^{2}-x_{i}^{2}|,\label{final3ux}\\
V(x,s)&=&\case{1}{2}Ns^{-1}x^{2}-\case{1}{4}\ln(x\sinh 2x),
\label{final3Vx}
\end{eqnarray}
\end{mathletters}%
with $\beta=2$ understood.

\subsection{Eigenvalue density}
\label{density}

The mean density $\langle\rho(x)\rangle_{s}$ of the $x$-variables is defined as
the ensemble average with distribution $P(\{x_{n}\},s)$ of the microscopic
density $\rho(x)$:
\begin{eqnarray}
&&\rho(x)=\sum_{n=1}^{N}\delta(x-x_{n}),\label{rhomicro}\\
&&\langle\rho(x)\rangle_{s}=\int_{0}^{\infty}\!\!dx_{1}
\int_{0}^{\infty}\!\!dx_{2}\cdots
\int_{0}^{\infty}\!\!dx_{N}\,P(\{x_{n}\},s)\rho(x).\label{rhomean}
\end{eqnarray}
The mean density is determined to leading order in $N$ by the integral equation
\begin{equation}
-\int_{0}^{\infty}\!\!dx'\,\langle\rho(x')\rangle_{s}\,
u(x,x')=V(x,s)+{\rm const}.\label{DYSONINT}
\end{equation}
The additive constant (which may depend on $s$ but is independent of $x$) is
fixed by the normalization condition
\begin{equation}
\int_{0}^{\infty}\!\!dx\,\langle\rho(x)\rangle_{s}=N.\label{normalizerho}
\end{equation}
Eq.\ (\ref{DYSONINT}) can be understood intuitively as the condition for
mechanical equilibrium of a fictitious one-dimensional gas with two-body
interaction $u$ in a confining potential $V$. Dyson\cite{Dys72} has shown that
corrections to Eq.\ (\ref{DYSONINT}) are an order $N^{-1}\ln N$ smaller than
the terms retained, and are $\beta$-dependent. These corrections are
responsible for the weak-localization effect in the conductance.\cite{Bee94}
Here we consider only the leading order contribution to the density, which is
of order $N$ and which is independent of $\beta$.

Substituting the functions $u(x,x')$ and $V(x,s)$ from Eq.\ (\ref{final3x})
into Eq.\ (\ref{DYSONINT}), and taking the derivative with respect to $x$ to
eliminate the additive constant, we obtain the equation
\begin{equation}
\frac{s}{2N}\int_{0}^{\infty}\!\!dx'\,\langle\rho(x')\rangle_{s}
\left(\frac{\sinh 2x}{\sinh^{2}x-\sinh^{2}x'}
+\frac{2x}{x^{2}-x'^{2}}\right)=x+{\cal O}(1/N).\label{Dysonint2}
\end{equation}
We note that
\begin{eqnarray}
\int_{0}^{\textstyle{s}}\!\!dx'
\left(\frac{\sinh 2x}{\sinh^{2}x-\sinh^{2}x'}
+\frac{2x}{x^{2}-x'^{2}}\right)&=&
\ln\left|\frac{\sinh(s+x)}{\sinh(s-x)}\right|
+\ln\left|\frac{s+x}{s-x}\right|\nonumber\\
&=&2x+{\cal O}(x/s),\;\;{\rm for}\;s\gg 1,\,s\gg x.\label{xovers}
\end{eqnarray}
It follows that the uniform density
\begin{equation}
\langle\rho(x)\rangle_{s}=\frac{N}{s}\,\theta(s-x)\label{rho0result}
\end{equation}
is the solution of Eq.\ (\ref{Dysonint2}) in the regime $s\gg 1$, $s\gg x$.
(The function $\theta(\xi)$ equals 1 for $\xi>0$ and 0 for $\xi<0$.) The result
(\ref{rho0result}) was first obtained by Mello and Pichard, by direct
integration of the DMPK equation.\cite{Mel89} To order $N$, the $x$-variables
have a uniform density of $Nl/L$, with a cutoff at $L/l$ such that the
normalization (\ref{normalizerho}) is satisfied. In the cutoff region $x\sim
L/l$ the density deviates from uniformity, but this region is irrelevant since
the transmission eigenvalues are exponentially small for $x\gg 1$.

\subsection{Correlation function}
\label{correlation}
The two-point correlation function $K(x,x',s)$ is defined by
\begin{equation}
K(x,x',s)=\langle\rho(x)\rangle_{s}\,\langle
\rho(x')\rangle_{s}-\langle\rho(x)\rho(x')\rangle_{s}.\label{twopoint}
\end{equation}
We compute the two-point correlation function by the general method of Ref.\
\onlinecite{Bee93}, which is based on an exact relationship between $K$ and the
functional derivative of the mean eigenvalue density $\langle\rho\rangle$ with
respect to the eigenvalue potential $V$:
\begin{equation}
K(x,x',s)=\frac{1}{\beta}\frac{\delta\langle\rho(x)\rangle_{s}}
{\delta V(x',s)}.\label{KVrelation}
\end{equation}
Eq.\ (\ref{KVrelation}) holds for any probability distribution of the form
(\ref{final3Px}), regardless of whether the interaction is logarithmic or not.
In the large-$N$ limit the functional derivative can be evaluated from the
integral equation (\ref{DYSONINT}). The functional derivative
$\delta\langle\rho\rangle/\delta V$ equals the solving kernel of
\begin{equation}
-\int_{0}^{\infty}\!\! dx'\,\psi(x')u(x,x')
=\phi(x)+{\rm const},\label{linearint2}
\end{equation}
where the additive constant has to be chosen such that $\psi$ has zero mean,
\begin{equation}
\int_{0}^{\infty}\!\! dx\,\psi(x)=0,\label{normalize2}
\end{equation}
since the variations in $\langle\rho\rangle$ have to occur at constant $N$.
Because of Eq.\ (\ref{KVrelation}), the integral solution
\begin{equation}
\psi(x)=\int_{0}^{\infty}\!\!dx'\,
\beta K(x,x')\phi(x')\label{solving}
\end{equation}
of Eq.\ (\ref{linearint2}) directly determines the two-point correlation
function. It turns out that $K(x,x')\equiv K(x,x',s)$ is independent of $s$ in
the metallic regime.

The integral equation (\ref{linearint2}) can be solved analytically by the
following method. We extend the functions $\psi$ and $\phi$ symmetrically to
negative $x$, by defining $\psi(-x)\equiv\psi(x)$, $\phi(-x)\equiv\phi(x)$. We
then note that the decomposition
\begin{mathletters}
\label{decompose}
\begin{eqnarray}
u(x,x')&=&-\case{1}{2}\,{\cal U}(x-x')-\case{1}{2}\,{\cal U}(x+x')+\ln
2,\label{decomposea}\\
{\cal U}(x)&=&\ln|2x\sinh x|,\label{decomposeb}
\end{eqnarray}
\end{mathletters}%
transforms the integral equation (\ref{linearint2}) into a convolution,
\begin{equation}
\int_{-\infty}^{\infty}\!\! dx'\,\psi(x')\,{\cal U}(x-x')
=2\phi(x)+{\rm const},\label{linearint3}
\end{equation}
which is readily solved by Fourier transformation. The Fourier transformed
kernel is
\begin{equation}
{\cal U}(k)\equiv\int_{-\infty}^{\infty}\!\!dx\,{\rm e}^{{\rm i}kx}\,
{\cal U}(x)=
-\frac{\pi}{|k|}[1+{\rm cotanh}\,(\case{1}{2}\pi |k|)].\label{Uk}
\end{equation}
The $k$-space solution to Eq.\ (\ref{linearint3}) is $\psi(k)=2\phi(k)/{\cal
U}(k)$, which automatically satisfies the normalization (\ref{normalize2}). In
$x$-space the solution becomes
\begin{mathletters}
\label{psixresult}
\begin{eqnarray}
\psi(x)&=&2\int_{-\infty}^{\infty}\!\!dx'\,{\cal K}(x-x')\phi(x')\nonumber\\
&=&2\int_{0}^{\infty}\!\!dx'\left[{\cal K}(x-x')+{\cal K}(x+x')\right]
\phi(x'),\label{psix}\\
{\cal K}(x)&=&\frac{1}{2\pi}\int_{-\infty}^{\infty}\!\!dk\,{\rm e}^{-{\rm
i}kx}\,\frac{1}{{\cal U}(k)}=
\frac{1}{\pi}\int_{0}^{\infty}\!\!dk\,\frac{\cos kx}{{\cal U}(k)}.
\label{calKx}
\end{eqnarray}
\end{mathletters}

Combining Eqs.\ (\ref{solving}), (\ref{Uk}), and (\ref{psixresult}), we find
that the two-point correlation function is given by
\begin{mathletters}
\label{Kresult}
\begin{eqnarray}
K(x,x')&=&{\cal K}(x-x')+{\cal K}(x+x'),\label{Kgdef}\\
{\cal K}(x)&=&-\frac{2}{\beta\pi^{2}}\int_{0}^{\infty}\!\!dk\,\frac{k\cos
kx}{1+{\rm cotanh}(\case{1}{2}\pi k)},\label{gdef}
\end{eqnarray}
with $\beta=2$. The inverse Fourier transform (\ref{gdef}) evaluates to
\begin{eqnarray}
{\cal K}(x)&=&\frac{1}{2\beta\pi^{2}}\,\frac{d^{\,2}}{dx^{2}}\ln[1+(\pi/x)^{2}]
\nonumber\\
&=&\frac{1}{\beta\pi^{2}}\,{\rm Re}\left[(x+{\rm i}0^{+})^{-2}-
(x+{\rm i}\pi)^{-2}\right],\label{Kxresult}
\end{eqnarray}
\end{mathletters}%
where $0^{+}$ is a positive infinitesimal.

We derived\cite{Rej94} these expressions for the two-point correlation function
for the case $\beta=2$. A direct integration of the DMPK equation by Chalker
and Mac\^{e}do\cite{Cha93} shows that the function $K(x,x')$ has in fact the
$1/\beta$ dependence indicated in Eq.\ (\ref{Kresult}), as expected from
general considerations.\cite{Bee93}

\subsection{Universal conductance fluctuations}
\label{UCFsection}
Now that we have the two-point correlation function, we can compute the
variance ${\rm Var}\,A=\langle A^{2}\rangle-\langle A\rangle^{2}$
of any linear statistic $A=\sum_{n=1}^{N}a(x_{n})$ on the transmission
eigenvalues (recall that $T_{n}\equiv\cosh^{-2}x_{n}$). By definition
\begin{equation}
{\rm Var}\,A=-\int_{0}^{\infty}\!\! dx\int_{0}^{\infty}\!\!
dx'\,a(x)a(x')K(x,x').\label{VarAKrelation}
\end{equation}
Substituting Eq.\ (\ref{Kresult}) we find
\begin{mathletters}
\label{VarAakx}
\begin{eqnarray}
&&{\rm
Var\,}A=\frac{1}{\beta\pi^{2}}\int_{0}^{\infty}\!\!dk\,
\frac{k|a(k)|^{2}}{1+{\rm cotanh}(\case{1}{2}\pi k)},\label{VarAresult}\\
&&a(k)=2\int_{0}^{\infty}\!\!dx\,a(x)\cos kx,\label{akdef}
\end{eqnarray}
or equivalently,
\begin{eqnarray}
{\rm Var}\,A=-\frac{1}{2\beta\pi^{2}}\int_{0}^{\infty}
\!\!dx\int_{0}^{\infty}
\!\!dx'\left(\frac{da(x)}{dx}\right)
\left(\frac{da(x')}{dx'}\right)\ln
\left(\frac{1+\pi^{2}(x+x')^{-2}}{1+\pi^{2}(x-x')^{-2}}\right).
\label{VarAresult2}
\end{eqnarray}
\end{mathletters}%

To obtain the variance of the conductance $G/G_{0}=\sum_{n}T_{n}$ (with
$G_{0}=2e^{2}/h$), we substitute $a(x)=\cosh^{-2}x$, hence $a(k)=\pi
k/\sinh(\frac{1}{2}\pi k)$, hence
\begin{equation}
{\rm Var\,}G/G_{0}=\frac{2}{15}\,\beta^{-1},\label{VarGresult}
\end{equation}
in agreement with Eq.\ (\ref{UCF}). In the same way one can compute the
variance of other transport properties. For example, for the shot-noise
power\cite{But90} $P/P_{0}=\sum_{n}T_{n}(1-T_{n})$ (with $P_{0}=2e|V|G_{0}$
and $V$ the applied voltage) we substitute $a(x)=\cosh^{-2}x-\cosh^{-4}x$,
hence $a(k)=\case{1}{6}\pi k(2-k^{2})(\sinh\case{1}{2}\pi k)^{-1}$, hence
\begin{equation}
{\rm Var\,}P/P_{0}=\frac{46}{2835}\,\beta^{-1},\label{VarPresult}
\end{equation}
in agreement with the result obtained by a moment expansion of the DMPK
equation.\cite{Jon92} Another example is the conductance $G_{\rm NS}$ of a
normal-superconductor junction, which for $\beta=1$ is a linear
statistic,\cite{Bee92} $G_{\rm NS}/G_{0}=\sum_{n}2T_{n}^{2}(2-T_{n})^{-2}$. We
substitute $a(x)=2\cosh^{-4} x\,(2-\cosh^{-2}x)^{2}=2\cosh^{-2}(2x)$, hence
$a(k)=\case{1}{2}\pi k/\sinh(\case{1}{4}\pi k)$, hence
\begin{equation}
{\rm Var\,}G_{\rm
NS}/G_{0}=\frac{16}{15}-\frac{48}{\pi^{4}}.\label{VarGNSresult}
\end{equation}
Finally, for the variance of the critical current $I_{\rm c}$ of a
point-contact Josephson junction (which is also a linear statistic for
$\beta=1$)\cite{Bee91,Shima} we compute
\begin{equation}
{\rm Var\,}I_{\rm c}/I_{0}=0.0890,\label{VarIcresult}
\end{equation}
with $I_{0}=e\Delta/\hbar$ and $\Delta$ the superconducting energy gap.

As in the previous subsection, we note that our results are derived for
$\beta=2$, and that the $1/\beta$ dependence of the variance in Eq.\
(\ref{VarAakx}) needs the justification provided by the calculation of Chalker
and Mac\^{e}do.\cite{Cha93}

\section{Insulating regime}
\label{insulating}

The solution (\ref{final}) can also be simplified in the regime $1\ll N\ll s$
of a conductor which is long compared to the localization length $Nl$. This is
the insulating regime. It is sufficient to consider the range $x_{n}\gg 1$,
since the probability that $x\lesssim 1$ is of order $N/s$ which is $\ll 1$.
The appropriate asympotic expansion of the Legendre function is
\begin{equation}
{\rm P}_{\frac{1}{2}({\rm i}k-1)}(\cosh 2x)=
(2\pi\sinh 2x)^{-1/2}\,{\rm Re}\,
\frac{2\Gamma(\frac{1}{2}{\rm i}k){\rm e}^{{\rm i}kx}}
{\Gamma(\frac{1}{2}+\frac{1}{2}{\rm i}k)}
\;\;{\rm for}\;x\gg 1.\label{Pxgg1expansion}
\end{equation}
For $s/N\gg 1$, the dominant contribution to the integral over $k$ in Eq.\
(\ref{final}) comes from the range $k\ll 1$. In this range
$\tanh(\case{1}{2}\pi k)\rightarrow\case{1}{2}\pi k$ and the ratio of Gamma
functions in Eq.\ (\ref{Pxgg1expansion}) simplifies to
\begin{equation}
\frac{\Gamma(\frac{1}{2}{\rm i}k)}
{\Gamma(\frac{1}{2}+\frac{1}{2}{\rm i}k)}=
(\case{1}{2}{\rm i}k\surd\pi)^{-1}
\;\;{\rm for}\;k\ll 1.\label{Gammaexpansion}
\end{equation}
The $k$-integration can now be carried out analytically,
\begin{equation}
\int_{0}^{\infty}\!\!dk\,\exp(-\case{1}{4}k^{2}s/N)
k^{2m-1}\sin(kx_{n})=(-1)^{m-1}\pi^{1/2}(N/s)^{m}
\exp(-x_{n}^{2}N/s)H_{2m-1}(x_{n}\sqrt{N/s}),\label{integralH}
\end{equation}
with $H_{2m-1}$ a Hermite polynomial. We then apply the determinantal identity
[cf.\ Eq.\ (\ref{Ldet})]
\begin{equation}
{\rm Det\,}H_{2m-1}(x_{n}\sqrt{N/s})=c\,{\rm
Det\,}x_{n}^{2m-1}=c\prod_{i}x_{i}\,\prod_{i<j}(x_{j}^{2}-x_{i}^{2}),
\label{Hdet}
\end{equation}
with $c$ an $x$-independent number. Collecting results, we find that the
general solution (\ref{final}) reduces in the insulating regime to
\begin{equation}
P(\{x_{n}\},s)=C(s)\prod_{i<j}\left[(\sinh^{2}x_{j}-\sinh^{2}x_{i})
(x_{j}^{2}-x_{i}^{2})\right]
\prod_{i}\left[\exp(-x_{i}^{2}N/s)x_{i}(\sinh
2x_{i})^{1/2}\right].\label{final2ins}
\end{equation}
(This formula was cited incorrectly in our Letter.\cite{Rej94})

The result (\ref{final2ins}) can be simplified further by ordering the
$x_{n}$'s from small to large and using that $1\ll x_{1}\ll x_{2}\ll\cdots\ll
x_{N}$ in the insulating regime ($s\gg N$). The distribution function then
factorizes,
\begin{eqnarray}
P(\{x_{n}\},s)&=&C(s)\prod_{i=1}^{N}
\exp\left[(2i-1)x_{i}-x_{i}^{2}N/s\right]
\nonumber\\
&=&(\pi s/N)^{-N/2}\prod_{i=1}^{N}
\exp\left[-(N/s)(x_{i}-\bar{x}_{i})^{2}\right].
\label{final3ins}
\end{eqnarray}
The $x_{n}$'s have a gaussian distribution with mean
$\bar{x}_{n}=\frac{1}{2}(s/N)(2n-1)$ and variance $\frac{1}{2}s/N$. The width
of the gaussian is smaller than the mean spacing by a factor $(N/s)^{1/2}$,
which is $\mbox{}\ll 1$, so that indeed $1\ll x_{1}\ll x_{2}\ll\cdots\ll
x_{N}$, as anticipated.

The conductance $G/G_{0}=\sum_{n}\cosh^{-2}x_{n}$ is dominated by $x_{1}$,
i.e.\ by the smallest of the $x_{n}$'s. Since $x_{1}\gg 1$ we may approximate
$G/G_{0}=4\exp(-2x_{1})$. It follows that the conductance has a log-normal
distribution, with mean $\langle\ln G/G_{0}\rangle=-s/N+{\cal O}(1)$ and
variance ${\rm Var}\,\ln G/G_{0}=2s/N$. Hence we conclude that
\begin{equation}
{\rm Var}\,\ln G/G_{0}=-2\langle\ln G/G_{0}\rangle,\label{VarlnG}
\end{equation}
in agreement with the result obtained by Pichard,\cite{Pic91} by directly
solving the DMPK equation in the localized regime.

The results obtained here are for the case $\beta=2$. Pichard has shown that
the relationship (\ref{VarlnG}) between mean and variance of $\ln G/G_{0}$
remains valid for other values of $\beta$, since both the mean and the variance
have a $1/\beta$ dependence on the symmetry index.

\section{Comparison with random-matrix theory}
\label{comparison}

The random-matrix theory of quantum transport\cite{Mut87,Sto91} is based on the
postulate that all correlations between the transmission eigenvalues are due to
the jacobian (\ref{jacobian}). The resulting distribution function
(\ref{Pglobal}) has the form of a Gibbs distribution with a logarithmic
repulsive interaction in the variables $\lambda_{n}\equiv (1-T_{n})/T_{n}$.
There exists a maximum-entropy argument for this
distribution,\cite{Mut87,Sto91} but it has no microscopic justification. In
this paper we have shown, for the case of a quasi-1D geometry without
time-reversal symmetry, that the prediction of RMT is highly accurate but not
exact.

In the {\em metallic regime\/} ($L\ll Nl$), the distribution is given by Eq.\
(\ref{final3x}). In terms of the $\lambda$-variables
($\lambda\equiv\sinh^{2}x$), the distribution takes the form (\ref{Pglobala})
of RMT, but with a different interaction
\begin{equation}
u(\lambda_{i},\lambda_{j})=-\case{1}{2}\ln|\lambda_{j}-\lambda_{i}|
-\case{1}{2}\ln|{\rm arcsinh}^{2}\lambda_{j}^{1/2}-{\rm
arcsinh}^{2}\lambda_{i}^{1/2}|.\label{ulambda1}
\end{equation}
For $\lambda\ll 1$ (i.e.\ for $T$ close to unity)
$u(\lambda_{i},\lambda_{j})\rightarrow -\ln|\lambda_{j}-\lambda_{i}|$, so we
derive the logarithmic eigenvalue repulsion (\ref{Pglobalb}) for the strongly
transmitting scattering channels. However, for $\lambda\approx 1$ the
interaction (\ref{ulambda1}) is {\em non-logarithmic}. For fixed
$\lambda_{i}\ll 1$, $u(\lambda_{i},\lambda_{j})$ as a function of $\lambda_{j}$
crosses over from $-\ln|\lambda_{j}-\lambda_{i}|$ to
$-\frac{1}{2}\ln|\lambda_{j}-\lambda_{i}|$ as $\lambda_{j}\rightarrow\infty$
(see Fig.\ 1). It is remarkable that, for weakly transmitting channels, the
interaction is twice as small as predicted by considerations based solely on
the jacobian. We have no intuitive argument for this result. The reduced level
repulsion for weakly transmitting channels enhances the variance of the
conductance fluctuations above the prediction (\ref{UCFglobal}) of RMT. Indeed,
as shown in Sec.\ III.D, a calculation along the lines of Ref.\
\onlinecite{Bee93}, but for the non-logarithmic interaction (\ref{ulambda1}),
resolves the $\case{1}{8}$~---~$\case{2}{15}$ discrepancy between RMT and
diagrammatic perturbation theory, discussed in the Introduction. The
discrepancy is so small because only the weakly transmitting channels (which
contribute little to the conductance) are affected by the non-logarithmic
interaction.

In the {\em insulating regime\/} ($L\gg Nl$), the distribution is given by Eq.\
(\ref{final2ins}). In terms of the $\lambda$'s the distribution takes the form
(\ref{Pglobala}) of RMT, but again with the non-logarithmic interaction
(\ref{ulambda1}). Since $\ln\lambda\gg 1$ in the insulating regime, the
interaction (\ref{ulambda1}) may be effectively simplified to
$u(\lambda_{i},\lambda_{j})= -\case{1}{2}\ln|\lambda_{j}-\lambda_{i}|$, which
is a factor of two smaller than the interaction (\ref{Pglobalb}) predicted by
RMT. This explains the factor-of-two discrepancy between the results of RMT and
of numerical simulations for the width of the log-normal distribution of the
conductance:\cite{Pic91} RMT predicts ${\rm Var}\,\ln G/G_{0}=-\langle\ln
G/G_{0}\rangle$, which is twice as small as the correct result (\ref{VarlnG}).

We conclude by mentioning some directions for future research. We have only
solved the case $\beta=2$ of broken time-reversal symmetry. In that case the
DMPK equation (\ref{DMPK}) can be mapped onto a free-fermion problem. For
$\beta=1,4$ the Sutherland-type mapping which we have considered is onto an
interacting Schr\"{o}dinger equation. It might be possible to solve this
equation exactly too, using techniques developed recently for the Sutherland
hamiltonian.\cite{Sim93,Muc93} From the work of Chalker and
Mac\^{e}do\cite{Cha93} we know that the two-point correlation function in the
large-$N$ limit has a simple $1/\beta$ dependence on the symmetry index. This
poses strong restrictions on a possible $\beta$-dependence of the eigenvalue
interaction.

Another technical challenge is to compute the exact two-point correlation
function $K(x,x',s)$ from the distribution function $P(\{x_{n}\},s)$. Our
result (\ref{final3x}) for $P$ is exact, but the large-$N$ asymptotic result
(\ref{Kresult}) for $K$ ignores fine structure on the scale of the eigenvalue
spacing. (This large-$N$ result for $K$ corresponds to the regime of validity
of the diagrammatic perturbation theory of UCF,\cite{Alt85,Lee85} while the
exact result for $P$ goes beyond perturbation theory.) In RMT there exists a
technique known as the method of orthogonal polynomials,\cite{Meh67} which
permits an exact computation of $K$.\cite{Mut93} A logarithmic interaction
seems essential for this method to work, and we see no obvious way to
generalize it to the non-logarithmic interaction (\ref{ulambda1}).

It might be possible to come up with another maximum-entropy principle,
different from that of Muttalib, Pichard, and Stone,\cite{Mut87} which yields
the correct eigenvalue interaction (\ref{ulambda1}) instead of the logarithmic
interaction (\ref{Pglobalb}). Slevin and Nagao\cite{Sle93} have recently
proposed an alternative maximum-entropy principle, but their distribution
function does not improve the agreement with Eq.\ (\ref{UCF}).\cite{Note2}

To go beyond quasi-one-dimensional geometries (long and narrow wires) remains
an outstanding problem. A numerical study of Slevin, Pichard, and
Muttalib\cite{Sle93b} has indicated a significant break-down of the logarithmic
repulsion for two- and three-dimensional geometries (squares and cubes). A
generalization of the DMPK equation (\ref{DMPK}) to higher dimensions has been
the subject of some recent investigations.\cite{Mel91b,Cha93b} It remains to be
seen whether the method developed here for Eq.\ (\ref{DMPK}) is of use for that
problem.

\acknowledgments
This research was supported in part by the ``Ne\-der\-land\-se
or\-ga\-ni\-sa\-tie voor We\-ten\-schap\-pe\-lijk On\-der\-zoek'' (NWO) and by
the ``Stich\-ting voor Fun\-da\-men\-teel On\-der\-zoek der Ma\-te\-rie''
(FOM).

\end{document}